%% file: Main.tex
\documentclass[journal,twoside,web]{ieeecolor}

\usepackage{generic}
\usepackage{cite}
\usepackage{amsmath,amssymb,amsfonts}
\usepackage{bbm}

\usepackage{amsthm}
\usepackage{array}
\usepackage{graphicx}
\usepackage{textcomp}
\usepackage{bbm}
\usepackage{glossaries}
\usepackage{tabularx}
\usepackage{multirow}
\usepackage{booktabs}

 \pdfoutput=1

\def\BibTeX{{\rm B\kern-.05em{\sc i\kern-.025em b}\kern-.08em
    T\kern-.1667em\lower.7ex\hbox{E}\kern-.125emX}}
\input{Input/Commands}

\DeclareMathOperator*{\argmin}{arg\,min}

\newcommand{\LL}[1]{\textcolor{magenta}{[LL: #1]}}

\newcommand{ \edit}[1]{\textcolor{black}{#1}}

\input{acronyms.tex}

\begin{document}

\title{
\edit{A Unifying Perspective for Safety of Stochastic Systems: From
Barrier Functions to Finite Abstractions}
}
\author{Luca Laurenti* and Morteza Lahijanian*
\thanks{Luca Laurenti is with the Delft University of Technology (e-mail: l.laurenti@tudelft.nl). }
\thanks{Morteza Lahijanian is with the University of Colorado Boulder (e-mail: morteza.lahijanian@colorado.edu).}
\thanks{* Equal contribution}}

\maketitle

\begin{abstract}

Providing safety guarantees for stochastic dynamical systems is a central problem in various fields, including control theory, machine learning, and robotics. Existing methods either employ \emph{Stochastic Barrier Functions} (SBFs) or rely on numerical approaches based on \emph{finite abstractions}. 
SBFs, analogous to Lyapunov functions, are used to establish (probabilistic) set invariance,
whereas abstraction-based approaches approximate the stochastic system with a finite model to compute safety probability bounds. 
This paper presents a \emph{unifying perspective} on these seemingly different approaches. Specifically, we show that both methods can be interpreted as approximations of a \emph{stochastic dynamic programming} problem.
\edit{
This perspective allows us to formally establish the correctness of both techniques, characterize their convergence and optimality properties, and analyze their respective assumptions, advantages, and limitations.
}
Our analysis reveals that\edit{, unlike SBFs-based methods, abstraction-based approaches can provide asymptotically optimal safety certificates, albeit at the cost of increased computational effort.}


\end{abstract}

\begin{IEEEkeywords}
 Finite Abstraction, Barrier Function,  Probabilistic Safety, Robustness,   Stochastic Systems 
\end{IEEEkeywords}

\section{Introduction}
    \label{sec:introduction}

\input{Input/Intro.tex}



\section{Probabilistic Safety for Discrete-Time Stochastic Processes}
    \label{sec:PropSafety}
    \input{Input/ProbForm.tex}

\section{Stochastic Barrier Functions}
    \label{sec:SBFs}
    \input{Input/SBF.tex}

\section{Discrete Abstraction}
    \label{sec:abstractions}
    \input{Input/Abstraction.tex}

\section{Abstractions vs Barriers: Pros and Cons}
 \label{sec:ProsAndCons}
 \input{Input/ProsandCons.tex}

\section{Conclusion and Future Directions}
\label{sec:Conclusion}


Safety analysis of stochastic systems is a major problem in today's world, especially as autonomous cyber-physical systems become increasingly integral.
Existing methods for such analysis are based on either stochastic barrier functions or discrete abstraction. Our analysis shows that both of these methods are approximations of the same stochastic dynamic programming problem. 
This perspective unifies these approaches and allows for a fair comparison, revealing that the methods are complementary, each with its own strengths and weaknesses. Consequently, the choice of approach should depend on the specific application under consideration.


Our demonstration of the effectiveness of these methods in solving the safety DP problem raises several open research questions. 
    The first question arises from the connection established between SBFs and abstraction-based methods and the fact that they are complementary. Specifically, given this knowledge, is it possible to devise new approaches that integrate their strengths, mitigate their weaknesses, and improve the scalability of existing methods?
    The second question stems from the observation that only a few approaches for controller synthesis algorithms exist, and they are generally limited to discrete or convex settings~\cite{jagtap2020formal,delimpaltadakis2022interval,santoyo2021barrier}.  Therefore, there is a need for more research to tackle the control synthesis problem with better and more general algorithms.   
    Another interesting question is related to what this paper does not consider.  That is, in this paper, we only consider systems with uncertainty rooted in the noise in the dynamics of the system (i.e., aleatoric uncertainty); however, often, the noise characteristics or the dynamics of the system are themselves uncertain. This setting has recently attractied interest from the control and machine learning communities, 
    with approaches based on abstractions to robust MDPs~\cite{gracia2023distributionally} and scenario-based optimization~\cite{badings2023probabilities,mathiesen2023inner}. 
    However, how to provide guarantees of safety and synthesize controllers for this more general setting, especially in the case where the noise is not additive, is still an open question. 


This work lays a theoretical foundation for computing safety guarantees for stochastic systems. By establishing a common ground for the development of both SBFs and abstraction-based methods, we hope to stimulate new research at the intersection of these frameworks. We believe these methods may eventually enable the attainment of safety guarantees for real-world control systems.

\section{Proofs}
    \label{sec:proofs}
    \input{Input/proofs}

\bibliographystyle{IEEEtran}
\bibliography{refs}

\begin{IEEEbiographynophoto}{Luca Laurenti} is an assistant professor at the Delft Center for Systems and Control at TU Delft and co-director of the  HERALD Delft AI Lab . He received his PhD from the Department of Computer Science, University of Oxford (UK), where he was a member of the Trinity College. Luca has a background in stochastic systems, control theory, formal methods, and artificial intelligence. His research work focuses on developing data-driven systems provably robust to interactions with a dynamic and uncertain world.
\end{IEEEbiographynophoto}

\begin{IEEEbiographynophoto}{Morteza Lahijanian} (Member, IEEE) is an assistant professor in the Aerospace Engineering Sciences department, an affiliated faculty at the Computer Science department, and the director of the Assured, Reliable, and Interactive Autonomous (ARIA) Systems group at the University of Colorado Boulder. He received a B.S. in Bioengineering at the University of California, Berkeley and a PhD in Mechanical Engineering at Boston University. He served as a postdoctoral scholar in Computer Science at Rice University. Prior to joining CU Boulder, he was a research scientist in the department of Computer Science at the University of Oxford. His awards include Ella Mae Lawrence R. Quarles Physical Science Achievement Award, Jack White Engineering Physics Award, and NSF GK-12 Fellowship. Dr. Lahijanian's research interests span the areas of control theory, stochastic hybrid systems, formal methods, machine learning, and game theory with applications in robotics, particularly, motion planning, strategy synthesis, model checking, and human-robot interaction. 
\end{IEEEbiographynophoto}

\end{document}

%% file: Input/Commands.tex
\newcommand{\reals}{\mathbb{R}}
\newcommand{\naturals}{\mathbb{N}}

\newcommand{\B}{\mathcal{B}}

\newcommand{\x}{\mathbf{x}}   

\renewcommand{\u}{\mathbf{u}}
\renewcommand{\v}{\mathbf{v}}
\newcommand\indicator{\mathbbm{1}}
\newcommand{\ind}[2]{\indicator(#2, #1)}

\newcommand{\safe}{\mathrm{s}}
\newcommand{\unsafe}{\mathrm{u}}
\newcommand{\reach}{\mathrm{reach}}

\newtheorem{theorem}{Theorem}

\newtheorem{proposition}{Proposition}
\newtheorem{corollary}{Corollary}
\newtheorem{lemma}{Lemma}
\newtheorem{definition}{Definition}
\newtheorem{remark}{Remark}

\newcommand{\expect}{\mathbb{E}}

\newcommand{\M}{\mathcal{M}}
\newcommand{\I}{\mathcal{I}}

\newcommand{\Pup}{\hat{P}}
\newcommand{\Plow}{\check{P}}

%% file: acronyms.tex
\newacronym{mdp}{MDP}{Markov Decision Process}
\newacronym{rss}{RSS}{Responsibility Sensitive Safety}
\newacronym{tpg}{TPG}{two-player game}
\newacronym{sos}{SoS}{Sum-of-Squares}
\newacronym{qcp}{QPC}{Quadratically Constrained Program}
\newacronym{sdp}{SDP}{Semi-definite Programming}
\newacronym{fcnn}{FCNN}{Fully-connected Neural Network}
\newacronym{bnn}{BNN}{Bayesian Neural Network}
\newacronym{mcmc}{MCMC}{Markov Chain Monte Carlo}
\newacronym{hmc}{HMC}{Hamiltonian Monte Carlo}
\newacronym{nuts}{NUTS}{No U-Turn Sampler}
\newacronym{dtss}{DT-SS}{Discrete-Time Stochastic System}
\newacronym{pac}{PAC}{Probably Approximately Correct}
\newacronym{psd}{PSD}{Positive Semi-Definite}
\newacronym{smt}{SMT}{Satisfiability Modulo Theory}
\newacronym{milp}{MILP}{Mixed-Integer Linear Programming}
\newacronym{lbp}{LBP}{Linear Bound Propagation}
\newacronym{ibp}{IBP}{Interval Bound Propagation}
\newacronym{nn}{NN}{Neural Network}
\newacronym{nbf}{NBF}{Neural Barrier Function}

%% file: Input/Intro.tex

In the age of autonomous systems, control systems have become ubiquitous, playing a pivotal role in \emph{safety-critical} applications. Examples span from autonomous vehicles~\cite{schwarting2018planning} to medical robots~\cite{dupont2021decade}, where the consequences of bad decisions are not only costly but can also prove fatal.
A common characteristic of these systems is the inherent complexity and nonlinearity of the dynamics that are subject to uncertainty due to physics (e.g., sensor or actuation noise) or algorithms (e.g., black-box controllers or perception).
Consequently, the question \emph{how to ensure the safety of stochastic control systems?} has emerged as a central research topic in various disciplines, including control theory, machine learning, formal methods, and robotics~\cite{adams2022formal,clark2021control,kushner1967stochastic}.

In stochastic systems, \emph{safety} generalizes the standard notion of \textit{stochastic stability} \cite{kushner1967stochastic}.  It is defined as the probability that the system avoids unsafe behavior, i.e., with high probability, the first exit time from a given safe set is greater than a specific threshold or  infinite. However, computing the exact safety probability is generally intractable, even in simple stochastic \emph{linear} systems. As a result, existing formal methods rely on \emph{under-approximations}, with two of the most commonly employed approaches being \emph{stochastic barrier functions} (SBFs)~\cite{prajna2007framework,clark2021control} and finite-abstraction methods~\cite{kushner2012stochastic,lavaei2022automated}. 
Similar to Lyapunov functions for stability, SBFs are energy-like functions that provide lower bounds on the probability of a stochastic system remaining within a safe set. On the other hand, abstraction-based methods abstract the original system into a finite-state stochastic process, typically a variant of a Markov chain, for for which efficient algorithms exist to compute safety probabilities~\cite{Lahijanian:CDC:2012,lahijanian2015formal,cauchi2019efficiency,dutreix2022abstraction}.


\edit{This paper presents a \emph{unifying perspective} on safety verification of discrete-time, continuous-space stochastic systems through the lens of of \emph{dynamic programming} (DP), bridging the gap between SBFs and abstraction-based methods.}
While traditionally seen as distinct, with SBFs typically derived from super-martingale conditions \cite{kushner1967stochastic}, 
\edit{we show that both approaches can be understood as approximations of the same DP framework.}
First, we establish that, for the class of systems considered in this paper, the safety probability  can be computed via DP, 
and that there always exists a deterministic Markov policy (strategy) that optimizes this probability.
We then show that both SBFs and abstraction-based methods can be seen as approximations of this DP framework. Specifically, for SBFs, we show that existing bounds can be obtained by over-approximating the indicator function of the unsafe set with a barrier function, i.e., a non-negative function that is at least one in the unsafe set.
On the other hand, 
abstraction-based methods, which convert the system into an (uncertain) Markov process, arise as piecewise-constant over- and under-approximations of the defined DP.

Viewing both SBF and abstraction-based methods as approximations  {of} a DP problem has several advantages: (i) it gives a unified treatment of safety for stochastic systems, (ii) it allows us to establish formal bounds and guarantees on the precision and correctness of these approaches, and (iii) it enables us to fairly compare these methods and highlight their strengths and weaknesses.  Specifically, we show  {how} abstraction-based methods can return tighter bounds on the safety probability compared to SBFs, and this generally comes at the cost of increased computational effort. 

 {
In summary, the main contributions of this paper are:
\begin{itemize}
    \item introducing a unifying perspective on SBF and abstraction-based methods as approximations of a stochastic dynamic programming algorithm,   
    \item providing formal proofs on the convergence of abstraction-based methods to the optimal safety probability and optimal strategies, as well as demonstrating the lack of convergence guarantees for 
    \edit{(Martingale-based) SBF methods to the optimal probability and strategies}, and
    \item discussing the pros and cons of each method, supported by theoretical proofs and empirical illustrations.
\end{itemize}
}


The paper is organized as follows. In Section \ref{sec:PropSafety}, we introduce the class of systems we consider and formally define probabilistic safety. In Section \ref{sec:SBFs}, we formally define SBFs. We present abstraction-based methods in Section \ref{sec:abstractions}.  Finally, in Section \ref{sec:ProsAndCons}, we analyze strengths and weaknesses of each method. Finally, we conclude the paper in Section \ref{sec:Conclusion}, highlighting important open research questions.  We provide all the proofs in Section~\ref{sec:proofs}.

%% file: Input/ProbForm.tex
Consider a discrete-time controlled  stochastic  system described by the following difference equation:
\begin{equation}
\label{eq:system_equation}
    \x[k+1] = F(\x[k],\u[k],\v[k]), 
\end{equation}
where $\x[k]$ is the state of the system at time $k \in \mathbb{N}$ taking values in $X\subseteq \reals^{n_x}$, and $\u[k]$ denotes the control or action at time $k$ taking values in compact set $U\subset \reals^{n_u}$. 
\edit{For a probability space $(\Omega,\mathcal{F},P_v)$, $\v[k]:\Omega \to V \subseteq \reals^{n_v} $ is a random variable with associated probability distribution $P_v$, which represents the noise affecting the system at every time step.}
\edit{$\v[k]$ is assumed to be independent and identically distributed at every time step $k$.}
Furthermore, sets $X$, $U$, and $V$ are all assumed to be appropriately (Borel) measurable sets. 
Finally, $F : X \times  U \times V \to X$ is a possibly non-linear measurable function representing the one-step dynamics  {of the system}.

\begin{remark}
\edit{
System~\eqref{eq:system_equation} represents a general model of a nonlinear controlled stochastic system, which includes a large class of stochastic models used in practice. For instance, in the case of additive noise, we have $F(\x[k],\u[k],\v[k])=\bar{F}(\x[k],\u[k])+\v[k]$. 
In this example, if $\bar{F} (\x[k], \u[k])$ is a neural network, then System (1) becomes a stochastic neural network dynamic model~\cite{adams2022formal,mazouzsafety}.
}
\end{remark}


\begin{definition}[Policy]
A \emph{feedback policy} (or strategy) $\pi=(\pi_0,\pi_1,...)$ for System~\eqref{eq:system_equation} is a sequence of universally measurable stochastic kernels such that, for every $k \geq 0$, $\pi_k:X^{k+1} \to \mathcal{P}(U)$\footnote{Note that we assume that $\pi_k$ is independent of the value of the previous actions, and only depends on previous states. This is without loss of generality because probabilistic safety, as defined in Definition \ref{def:SafetyFiniteTime}, only depends on the state values.}, where $\mathcal{P}(U)$ is the set of probability measures over $U$. 
  {Policy} $\pi$ is called \emph{deterministic} if for each $k$ and $(x_0,...,x_k)$, $\pi_k(x_0,...,x_k)$ assigns probability mass one to some $u\in U$. In this case, with a slight abuse of notation, $\pi=(\pi_0,\pi_1,...)$ can be considered as the sequence of universally measurable functions $\pi_k:X^{k+1} \to U.$
If for each $k$, $\pi_k$  {only depends on} $x_k,$ $\pi$ is a \emph{Markov policy}. 
A policy is \emph{stationary} if, for every $k_1,k_2 \in \mathbb{N}$, it holds that $\pi_{k_1}=\pi_{k_2},$ in which case, with an abuse of notation, we use $\pi$ to denote any of these functions. The set of all policies is denoted by $\Pi,$ while the set of deterministic Markov policies by $\Pi^{M,D}$.
\end{definition}





For a given initial condition $x_0,$ a time horizon $H \in \naturals$, and a policy $\pi=(\pi_0,...,\pi_{H-1})$, ${\mathbf{{x}}}[k]$ is a stochastic process with a well defined probability measure $P$ generated 
by the noise distribution $p_{\v_k}$
\cite[Proposition 7.45]{bertsekas2004stochastic} such that for \edit{$k\in\{0,...,H-1\}$} and measurable sets $X_0, X_{k+1} \subseteq X$, 
it holds that
\begin{equation*}
    \label{Eqn:ProbDefinition}
    \begin{split}
        &P({\x}[0] \in X_0\mid \u[0]=a) = \ind{X_0}{x_0},\\
        &P({\x}[k + 1] \in X_{k+1} \mid  {\x}[k] = x, \u_k=a)\\
         &\hspace{3cm} := \int_{\Omega} \edit{\ind{X_{k+1}}{F(x,\edit{\v[k](\omega)},a)}\edit{P_v(d \omega)}} \\
        &\hspace{3cm} := T(X_{k+1} \mid  x, a ),
    \end{split}
\end{equation*}
where 
\begin{equation*}
    \ind{X_k}{x_k}=
    \begin{cases}
        1 & \text{if }x_k \in X_k\\
        0 & \text{otherwise}
    \end{cases}    
\end{equation*}
is the indicator function. 
We refer to $T(X_{k+1} \mid x_k,a)$ as the \textit{stochastic kernel} of System~\eqref{eq:system_equation}, and we assume that for each measurable $X_j\subseteq X,$ $x\in X,a\in U,\, T(X_j\mid x,a)$ is Lipschitz continuous in both $x$ and $a$.

\subsection{Probabilistic Safety}
For a given policy $\pi$ and a time horizon $H\in \naturals$, probabilistic safety is defined as the probability that ${\x}[k]$ stays within a  measurable safe set $X_\safe \subseteq X$ for the next $H$ time steps, i.e., the first exit time from $X_s$ is greater than $H.$\footnote{\edit{Depending on the application domain, horizon $H$ can be 
the prediction horizon (e.g., in MPC) or long-term time horizon (e.g., in motion planning).}
}

\begin{definition}[Probabilistic Safety]
    \label{def:SafetyFiniteTime}
    Given a policy $\pi,$ safe set $X_\safe \subset X$, time horizon $H\in \naturals$,  and initial set of states $ X_0 \subseteq X_\safe$, probabilistic safety is defined as
    \begin{multline*}
        P_{\safe}(X_\safe,X_0,H\mid \pi)= \\
        \inf_{x_0\in X_0}P(\forall k\in [0,H], {{\x}}[k] \in X_\safe \mid \x[0]=x_0,\pi).
    \end{multline*}
\end{definition}

Probabilistic safety, and its equivalent dual, probabilistic reachability\footnote{Given a finite-time horizon $H \in \naturals$, policy $\pi$, initial point $x_0$, and a target set $X_\unsafe$, probabilistic reachability is defined as $P_{\reach}(X_\unsafe,x_0,H \mid \pi) = P(\exists k \in [0,H], \x[k] \in X_\unsafe \mid \x[0]=x_0,\pi).$ Consequently, we have that $P_{\safe}(X_\safe,x_0,H \mid \pi)=1-P_{\reach}(X\setminus X_\safe,x_0,H \mid \pi)$. } are widely used to certify the safety of dynamical systems \cite{abate2008probabilistic} and represent a generalization of the notion of invariance that is commonly employed for analysis of deterministic systems~\cite{ames2016control}. 

\begin{remark}

    
    In Definition~\ref{def:SafetyFiniteTime}, we consider only a finite horizon $ H $, but this is without loss of generality. For an infinite horizon, one of two cases occurs:  
    \begin{itemize}
        \item There exists a sub-region of $ X_\safe $ from which the system never exits, in which case $ P_{\safe} $ corresponds to the probability of eventually reaching this region while remaining safe.  
        \item Otherwise, $ P_{\safe} = 0 $, as in the case where $ \mathbf{v} $ is additive and its distribution has unbounded support (e.g., a Gaussian distribution).  
    \end{itemize}
\end{remark}

In the remainder of this section, we show that, to compute a policy that maximizes probabilistic safety, it is sufficient to restrict to deterministic Markov policies.
To that end, we first show that $P_{\safe}$ can be characterized as the solution of a DP problem. 
In particular, for $k\in \{H, H-1, \ldots ,0 \}$ and $X_\unsafe := X\setminus X_\safe$, consider value functions $V^*_k:X \to [0,1]$ defined  recursively (backwardly in time) as:
\begin{align}
    \label{Eq:OptimalValueIterationAction}
    &V^{*}_H(x)=\ind{X_\unsafe}{x},\\
    &V^{*}_{k}(x)=\inf_{a\in U} \Big( \ind{X_\unsafe}{x} +  \ind{X_\safe}{x} \, \expect_{x' \sim T(\cdot \mid x, a)}[ V^{*}_{k+1}(x')   ]\Big),
\label{eqn:valueIterationAction}
\end{align}
where notation $ {x' \sim T(\cdot \mid x, a)}$ means that $x'$ is distributed according to $T(\cdot \mid x, a),$ \edit{and $ \expect_{x' \sim T(\cdot \mid x, a)}[ V^{*}_{k+1}(x')   ]$ is the expectation of value function $V^{*}_{k+1}$ with respect to $T(\cdot \mid x, a)$}.
Intuitively, at each time step $k$, $V^{*}_k$ selects the (deterministic) action that minimizes the probability of reaching a state from which the system may reach $X_\unsafe$ in the next $H-k$ time steps. Consequently, by propagating  $V^{*}_k$ backward over time, we compute the probability of reaching $X_\unsafe$ in the future.
The following theorem guarantees that
$\sup_{\pi \in \Pi}P_{\safe}(X_\safe,x_0,H \mid \pi)$ is equal to $1 - V^{*}_{0}(x_0)$.

\begin{theorem}
    \label{th:OptimalityMarkovPolicies}
    For an initial state $x_0\in X_\safe$, it holds that
    $$ \sup_{\pi \in \Pi}P_{\safe}(X_\safe,x_0,H\mid \pi)= 1 - V^{*}_0(x_0).$$
\end{theorem}
 {The proof is in Section \ref{sec:proofs},} \edit{where we show that standard results for optimality of deterministic policies commonly derived for cumulative or discounted reward models \cite{bertsekas2004stochastic} extend to probabilistic safety. }
Therefore, a straightforward consequence of Theorem~\ref{th:OptimalityMarkovPolicies} is  Corollary~\ref{Cor:DeterministicBellman}, which guarantees that deterministic Markov policies are optimal. In fact, note that in obtaining $V_0^*$ via \eqref{Eq:OptimalValueIterationAction}-\eqref{eqn:valueIterationAction}, we only consider deterministic Markov policies.

\begin{corollary} 
    \label{Cor:DeterministicBellman}
    \edit{Deterministic Markov policies are sufficient for optimal probabilistic safety, i.e, it hold that}
        $$ 
        \sup_{\pi \in \Pi} P_{\safe}(X_\safe,x_0,H\mid \pi) = \sup_{\pi \in \Pi^{M,D}} P_{\safe}(X_\safe,x_0,H\mid \pi).  
        $$
     Furthermore, for every $\pi \in \Pi^{M,D}$, it holds that   
     $$P_{\safe}(X_\safe,x_0,H\mid \pi)=1-V_{0}^{\pi}(x_0),$$ 
     where  $V_{0}^{\pi}(x_0)$ is defined recursively as 
    \begin{align}
        \label{Eq:FirstValueIteration}
        &V^{\pi}_H(x)=\ind{X_\unsafe}{x},\\
        &V_{k}^{\pi}(x)=\ind{X_\unsafe}{x} +  \ind{X_\safe}{x} \, \expect_{x' \sim T(\cdot \mid x, \pi_k(x))}[ V^{\pi}_{k+1}(x')].
        \label{eqn:valueIteration}
    \end{align}
\end{corollary}

Theorem \ref{th:OptimalityMarkovPolicies} and Corollary \ref{Cor:DeterministicBellman} guarantee that in order to synthesize optimal policies, it is enough to consider deterministic Markov policies, and these policies can be computed via DP. 

We should stress that without the assumptions made in this paper (compactness of $U$, continuity of $T$, and measurability of the various sets), $V^*_k$ may not be measurable and the infimum in $U$ in \eqref{eqn:valueIterationAction} may not be attained. 
In that case, the integrals in the expectations in  \eqref{eqn:valueIterationAction} and \eqref{eqn:valueIteration} 
have to be intended as outer integrals  \cite{bertsekas2004stochastic}.
However, under the assumptions in this paper,  the expectations in the above DP are well-defined and a universally measurable deterministic Markov optimal policy exists~\cite{bertsekas2004stochastic}, i.e., the $\inf$ is attainable for every point in the state space by a universally measurable function.
\edit{Nevertheless, even if $V^{*}_0$ and $V^{\pi}_0$ are well-defined, computation of \eqref{eqn:valueIterationAction} is infeasible in practice due to the need to solve uncountably many optimization problems. Thus, computation for $P_{\safe}$ requires approximations. }

In what follows, we consider the two dominant approaches in the literature that (with certified error bounds) compute 
probabilistic safety and synthesize policies for System~\eqref{eq:system_equation}, namely,
\emph{stochastic barrier functions} (SBFs) and \emph{abstraction-based methods}. We show that both approaches arise as over-approximations of the value functions $V^{\pi}_k$. 
 {
In other words,
both SBF-based and abstraction-based methods are approximation techniques to solve the same DP in \eqref{Eq:FirstValueIteration}-\eqref{eqn:valueIteration}.
}
Such a unified framework 
provides a basis for a fair comparison of these approaches, which consequently reveals their advantages and disadvantages. \edit{We start by considering SBF in Section~\ref{sec:SBFs} and abstraction-based methods in Section~\ref{sec:abstractions}.}


%% file: Input/SBF.tex
We begin with the setting, where a deterministic Markov
policy $\pi$ 
is given, and we aim to compute a (non-trivial) lower bound of $P_{\safe}(X_\safe,X_0,H\mid {\pi})$.  We first show how one can use SBFs to compute an upper bound on $V_0^{\pi}$ (hence, a lower bound on $P_\safe$) without the need to directly evolve the dynamics of System \eqref{eq:system_equation} over time. Then, we focus on the key challenge with SBFs: how to find an SBF that allows one to bound $P_\safe$ without leading to overly conservative results. The control synthesis case is considered in Section~\ref{Sec:ControlSBF}.

An SBF~\cite{prajna2007framework,santoyo2021barrier} is simply a continuous almost everywhere function $B: \reals^{n_x} \to \reals_{\geq 0} $ that over-approximates $V^{\pi}_H(x)$\edit{, that is, the indicator function for the unsafe set}. In particular, we say function $B$ is a SBF 
iff
\begin{align}
    \label{Eqn:BarrierConditions}
      \forall x\in X,  \,  B(x)\geq 0 \quad \text{and} \quad   \forall x\in X_\unsafe, \,   B(x)\geq 1. 
\end{align}
The intuition is that, when $B$ is propagated backwards over time in a DP fashion, it produces an over-approximation for $V^{\pi}_k(x)$.
That is, value functions $\bar{V}^{\pi}_k:\reals^{n_x} \to [0,1]$ with $k\in \{0, \ldots, H \}$ defined recursively as 
\begin{align}
    \label{eqn:valueIterationBarrierInitial}
    &\bar{V}^{\pi}_H(x)=B(x),\\
    &\bar{V}_{k}^{\pi}(x)=\ind{X_\unsafe}{x} B(x)  + \nonumber \\
    & \hspace{27mm} \ind{X_\safe}{x} \, \expect_{x' \sim T(\cdot \mid x, \pi_k(x))} [ \bar{V}^{\pi}_{k+1}(x')],
    \label{eqn:valueIterationBarrier}
\end{align} 
over-approximate $V^{\pi}_k(x)$, as formalized in the following lemma. 
\begin{lemma}
    \label{lemma:BarrierBoundValueFunction}
    Consider the value functions $V^{\pi}_k(x)$ and $\bar{V}_{k}$ defined in \eqref{Eq:FirstValueIteration}-\eqref{eqn:valueIteration} and \eqref{eqn:valueIterationBarrierInitial}-\eqref{eqn:valueIterationBarrier}, respectively.
    For every $k\in \{0, \ldots ,H \}$ and every $x\in X$, it holds that $V^{\pi}_k(x) \leq \bar{V}^{\pi}_k(x).$
\end{lemma}

 {Note that a sufficient condition for the inequality in Lemma~\ref{lemma:BarrierBoundValueFunction} to become a strict inequality is that $\exists x \in X_s$ such that $B(x)>0$ and $T(\cdot \mid x, \pi_k(x))$ has unbounded support for every $x$. In fact, this guarantees that for any $x\in X_s,$  $\expect_{x' \sim T(\cdot \mid x, \pi_k(x))} [ B(x')]>\expect_{x' \sim T(\cdot \mid x, \pi_k(x))} [ \ind{X_\unsafe}{x'}].$ }

\edit{To obtain $\bar{V}_{k}^{\pi},$ one needs to compute the expectation in \eqref{eqn:valueIterationBarrier} for uncountably many $x,$ which is generally intractable.} To overcome this problem, \edit{Martingale-based} SBFs, define
constant $\beta\geq 0$ as 
\begin{align}
\label{Eqn:BetaTerm}
\beta \geq  \sup_{x\in X_\safe, k \in \{0,...,H-1\}} \Big(\expect_{x' \sim T( \cdot \mid x, \pi_k(x))}[ B(x')   ]  - B(x) \Big).
\end{align}
That is, $\beta$ bounds how much the probability of reaching $X_\unsafe$ can grow in a single time step. Then, by rearranging terms in \eqref{eqn:valueIterationBarrier}, we obtain
\begin{align}
{V}_{k}^{\pi}(x) \; \leq   \; \bar{V}_{k}^{\pi}(x) \; \leq  \; (H-k)\beta + B(x).
    \label{eqn:valueIterationBarrierBeta}
\end{align}
This leads to the following theorem.
\begin{theorem}[\!\!{\cite{santoyo2021barrier}}\,]
\label{th:StochasticBarrierFunction}
Let $ B:X \to \reals_{\geq 0} $ be a function that satisfies~\eqref{Eqn:BarrierConditions}, and consider the bounds
$\eta = \sup_{x\in X_0} B(x)$ and $\beta$ in \eqref{Eqn:BetaTerm}.
Then, it holds that 
$$ \inf_{x_0 \in X_0} P_\safe^{}(X_\safe,x_0,H\mid {\pi}) \geq 1 - (\eta + \beta H).$$
\end{theorem}
\noindent

\edit{
Theorem \ref{th:StochasticBarrierFunction} ensures that once the constants $ \eta $ and $ \beta $ are determined, one can establish a lower bound on $ P_\safe^{}(X_\safe, x_0, H \mid \pi) $ for all initial states $x_0 \in X_0$. As a result, Theorem \ref{th:StochasticBarrierFunction} reduces solving the DP in \eqref{eqn:valueIterationAction} to quantifying the constants $ \eta $ and $ \beta $.
}

\begin{figure}[t]
    \centering
    \includegraphics[width=0.99\linewidth]{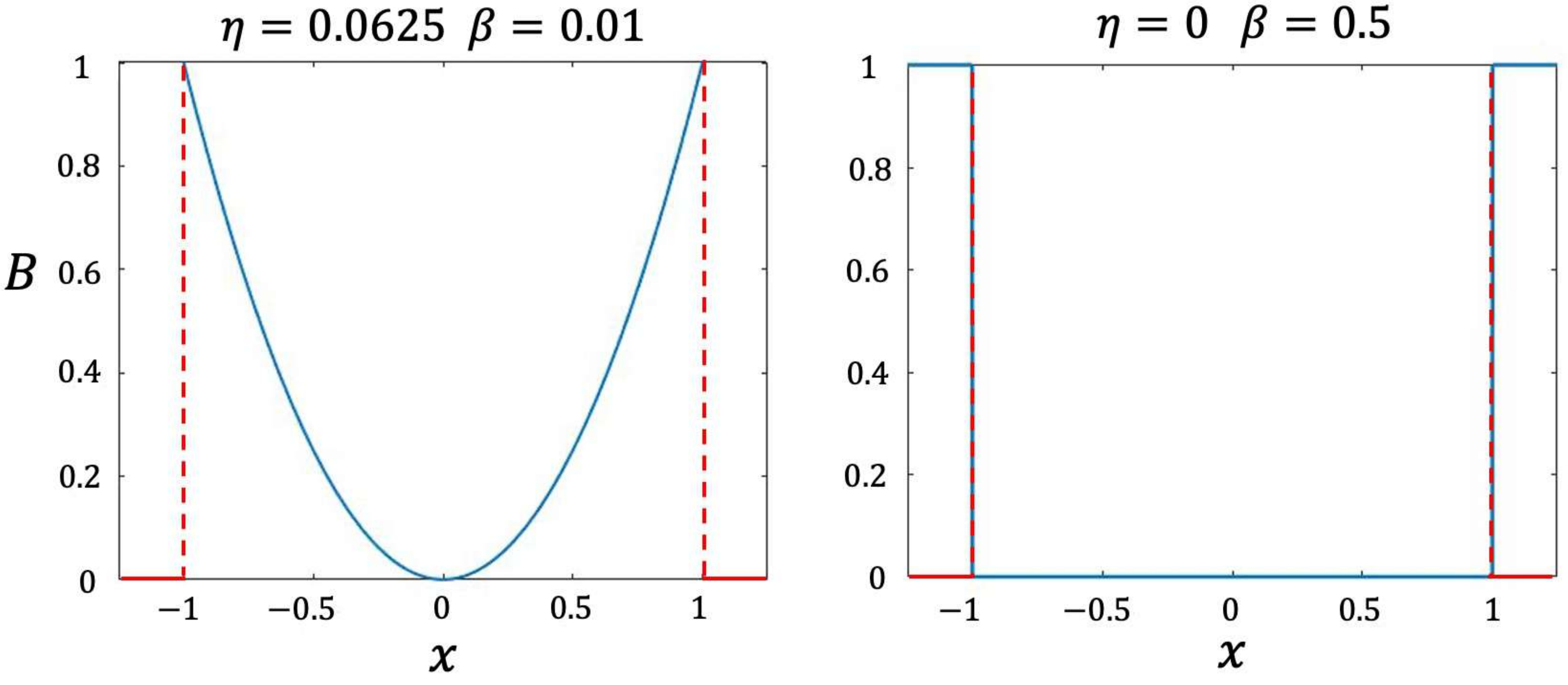}
    \caption{
    For a stochastic system $\mathbf{x}[k+1]=\mathbf{x}[k] + \mathbf{v}$, where $\mathbf{v}\sim \mathcal{N}(0,0.1)$ is an independent additive zero-mean Gaussian noise with standard deviation $0.1$. We consider $X_\safe=[-1,1]$ and $X_0=[-0.25,0.25]$. Two different stochastic barrier functions are shown: (left) a 4th degree polynomial as synthesized via SoS optimization for a time horizon $H=10$  { using the approach in \cite{santoyo2021barrier}, and (right) the indicator function on the unsafe set  {for which $\beta=\int_{(\infty,-1]\cup [1,\infty]}\mathcal{N}(x\mid 1,0.1)dx$}.    
    We obtain a lower bound on $P_\safe$ of $0.8275$ for SoS (left) and $0$ for the indicator function (right).
    }
    }
    \label{fig:ex_1_and_3}
\end{figure}

\begin{remark}
    \label{remark:SBFs}
    From \eqref{eqn:valueIterationBarrierInitial} and \eqref{eqn:valueIterationBarrier},
    we can observe that the closer $B(x)$ is to the indicator function, the closer $\bar{V}_{k}^{\pi}$ gets to $V_{k}^{\pi}$, and thus the tighter the bound computation becomes for $P_\safe$. This may lead the reader to wonder why we do not simply assume $B(x)=\ind{X_\unsafe}{x}.$ To clarify this, note  that, in the derivation of Theorem~\ref{th:StochasticBarrierFunction}, another source of conservatism comes from the choice of $\beta$. In fact, $\beta$ is the supremum expected change over all $x\in X_\safe$. Hence, setting $B(x)=\ind{X_\unsafe}{x}$ may lead to overly conservative results, e.g., cases where there are only few regions from which the probability that the system transition to the unsafe set is not negligible. This is illustrated in \edit{Fig.} \ref{fig:ex_1_and_3}, where an indicator barrier function is compared against a SBF synthesized using Sum-of-Square (SoS) optimization as proposed in \cite{santoyo2021barrier}. This example illustrates  {how the choice of $B$ can have a \edit{large} impact on $\beta$ and $\eta$, and consequently, on the resulting bounds on $P_\safe$.}
\end{remark}

\edit{Following Remark~\ref{remark:SBFs}, it is clear that the key challenge in the SBF approach is finding a valid $ B $ that avoids excessively conservative results while also allowing for the efficient computation of $ \eta $ and $ \beta $.}
Let $\B\subset \{f:\reals^{n_x}\to \reals_{\geq 0}\}$ be a class of non-negative functions, e.g., exponential or  {SoS}. Then,  the problem of searching for a valid barrier can be formulated as the solution of the following optimization problem:
\begin{align}
    \label{Eqn:SynthesisBarrierNoControl}   
    & \max_{B \in \B} \; 1-(\eta + \beta H) \\
    \nonumber    & \text{ subject to:} \\
    \nonumber    & \qquad \qquad    \inf_{x\in X_\unsafe} B(x)\geq 1,   \\
    \nonumber    & \qquad \qquad    \eta = \sup_{x\in X_0}B(x), \\
    \nonumber    & \qquad \qquad    \beta = \sup_{x\in X_\safe} \big(\expect_{x' \sim T(x'\mid x, \pi(x))}[ B(x')   ]  - B(x) \big). 
\end{align}

In the case where $\B$ is the class of exponential 
or SoS functions and dynamics function $F$ is polynomial in $\x$ and linear in $\v$, the above optimization problem can be reformulated as  a convex optimization problem \cite{santoyo2021barrier,steinhardt2012finite}. However, in the more general setting, the optimization problem in \eqref{Eqn:SynthesisBarrierNoControl} is non-convex, 
requiring relaxations that typically lead to partitioning of $X_s$, or ad-hoc methods to solve it efficiently \cite{mathiesen2022safety,mazouzsafety,abate2021learning}.

 {
\begin{remark}
    \label{remark: SBF approximation}    
    The solution to the optimization problem in \eqref{Eqn:SynthesisBarrierNoControl} is also a solution to the DP in \eqref{eqn:valueIterationBarrierInitial}-\eqref{eqn:valueIterationBarrier}; hence, the obtained solution is an over-approximation of the solution for the unifying DP in \eqref{Eq:FirstValueIteration}-\eqref{eqn:valueIteration}. This simple formulation and its computational tractability are primarily enabled by the conditions on $\eta$ defined in Theorem~\ref{th:StochasticBarrierFunction} and $\beta$ in \eqref{Eqn:BetaTerm}. However, these same $\eta$ and $\beta$ conditions are also the main sources of conservatism in lower-bounding $P_\safe$ in SBF methods.
\end{remark}
}

\begin{remark}
Note that  Theorem \ref{th:StochasticBarrierFunction} only requires $B$ to be almost surely continuous. That is, $B$ can be discontinuous in a set of (Lebesque) measure $0$. Consequently, $B$ can be taken to be a piecewise continuous functions. In practice, this can often be advantageous especially in the cases where $F$ is nonlinear~\cite{mathiesen2023inner}.  
\end{remark}

\subsection{Control Synthesis with SBFs}
\label{Sec:ControlSBF}

We now study how to find a policy $\pi$ that maximizes $P_\safe$ using SBFs. In particular, consider a stationary policy (feedback controller) $\pi(\cdot \mid \theta): X \to U$ parameterized in some parameters $\theta \in \reals^{n_\theta}$. Then, control synthesis with SBFs can be performed by modifying the optimization problem in~\eqref{Eqn:SynthesisBarrierNoControl} to the following:
\begin{align}
    \label{Eqn:SynthesisBarrier}   
    &\max_{B \in \B,\theta \in \reals^{n_\theta}} 1-(\eta + \beta H) \\
    \nonumber   &\text{ subject to:} \\
    \nonumber    & \qquad \qquad    \inf_{x\in X_\unsafe} B(x)\geq 1,   \\
    \nonumber    & \qquad \qquad    \eta = \sup_{x\in X_0}B(x), \\
    \nonumber    & \qquad \qquad    \beta = \sup_{x\in X_\safe} \big(\expect_{x' \sim T(x'\mid x, \pi(x\mid \theta))}[ B(x')   ]  - B(x) \big). 
\end{align}
This optimization problem aims to simultaneously synthesize a barrier function $B$ and a stationary policy (feedback controller)  $\pi$. Unfortunately, because the expectation in the $\beta$ (third) constraint depends on both $\pi$ and $B$, the resulting optimization problem is generally non-convex~\cite{agrawal2017discrete,jagtap2020formal}. To address this problem, recent approaches employ iterative methods~\cite{mazouzsafety,prajna2007framework,lavaei2022automated}. 
They generally proceed by first finding a $B$ for a fixed $\pi$ and then updating $\pi$ to maximize the lower bound on $P_\safe$, i.e., $1-(\eta + \beta H)$, for the fixed $B$.  
Other recent methods include employing machine learning algorithms~\cite{vzikelic2023learning}.

As discussed above, there are two main sources of conservatism in bounding $P_\safe$ using SBFs: (i) the choice of the barrier, and  (ii) the $\beta$ term in \eqref{Eqn:BetaTerm}, which is obtained as a uniform bound over the safe set.  Both of these sources of errors can be mitigated by abstraction-based approaches, usually at the price of increased computational effort. 

%% file: Input/Abstraction.tex


Another class of well-established approaches to compute $P_\safe$ and a policy $\pi$ that maximizes $P_\safe$ is based on finite abstraction~\cite{lavaei2022automated}.  These approaches aim to 
numerically solve the continuous 
DP in \eqref{Eq:FirstValueIteration}-\eqref{eqn:valueIteration} by discretizing $X$, while accounting for the error induced by discretization. We again start with the case where a deterministic Markov policy $\pi$ is given, and consider the control synthesis case in Section \ref{sec:ControlAbstractions}.

Abstraction-based methods first partition the safe set $X_\safe$ into  $n_p$ sets $X_1, \ldots ,X_{n_p}$ and treat $X_\unsafe$ as another discrete region (i.e., $X_{n_p+1}=X_\unsafe$), resulting in a total of $n_p+1$ regions.
For $k\in\{0,\ldots,H-1\}$ and a given policy $\pi$, one can then define piecewise-constant functions 
\begin{equation*}
    \gamma^{\pi}_{k}(x) = 
    \begin{cases}
        \gamma^\pi_{k,1} & \text{if } x \in X_1\\
        \; \; \vdots & \quad \; \vdots \\
        \gamma^\pi_{k,n_p} & \text{if } x \in X_{n_p}\\
        \gamma^\pi_{k,n_{p}+1} & \text{otherwise}
    \end{cases}
\end{equation*}
recursively, for $i\in\{1,\ldots, n_{p}+1 \}$, as\footnote{Note that under the assumption that $\pi$ and $T$ are locally continuous in $X_i$, we can replace the supremum with the maximum in the DP in \eqref{eqn:StartingConditionValueIt}-\eqref{eqn:RecursiveConditionValueIt}.  In fact,
a continuous function on a compact set attains its maximum and minimum on this set.}: 
\begin{align}
    \label{eqn:StartingConditionValueIt}    
    & \gamma^{\pi}_H(x)=\gamma^{\pi}_{H,i}:=\max_{ {\bar{x}} \in X_i} \ind{X_\unsafe}{\bar{x}},\\
    \nonumber   & \gamma^{\pi}_{k}(x)=\gamma^{\pi}_{k,i}:=\max_{ {\bar{x}} \in X_i} \Big( \ind{X_\unsafe}{ {\bar{x}}}  +\\
    & \qquad \qquad \qquad \qquad \ind{X_\safe}{ {\bar{x}}} \sum_{j=1}^{n_p+1} \gamma^{\pi}_{k+1,j} T(X_j\mid  {\bar{x}},\pi_k( {\bar{x}})) \Big).
    \label{eqn:RecursiveConditionValueIt} 
\end{align}
\edit{Intuitively, each $\gamma^{\pi}_{k,i}$ provides an upper-bound on the worst-case probability that, starting from any point in $X_i$ at time $k$, System \eqref{eq:system_equation} will reach the unsafe region before time $H$.  
Consequently, due to the maximum operator in \eqref{eqn:StartingConditionValueIt}-\eqref{eqn:RecursiveConditionValueIt}, $\gamma^{\pi}_k(x)$ over-approximates $V^{\pi}_{k}(x)$, which is the probability of reaching the unsafe set starting from $x$ in the next $H-k$ time steps. This leads to the following theorem.
}
\begin{theorem}
\label{th:CorrectnessGammaBoundsFixedControl}
    Let $\gamma^{\pi}_{k}$ be defined as in \eqref{eqn:StartingConditionValueIt}-\eqref{eqn:RecursiveConditionValueIt}. Then, it holds that 
    $$ P_\safe(X_\safe,x_0,H\mid \pi) \geq 1- \gamma^{\pi}_0(x_0). $$
\end{theorem}

\begin{proof}
    Because of Theorem~\ref{th:OptimalityMarkovPolicies} it is enough to show that for each $k\in\{0,...,H\}$ and $x\in X$, it holds that  $ V_k(x) \leq \gamma^{\pi}_k(x) .$ The case $x\in X_i$ for $X_i \cap X_\unsafe \neq \emptyset$ is trivially verified. Consequently, in what follows we assume $x\in X_i\subseteq X_\safe$. The proof is by induction. Base case is $k=H-1$, in this case we have
    \begin{align*}
        V^{\pi}_{H-1}(x)&=T(X_\unsafe \mid x,\pi_{H-1}(x))\\
        &\leq \max_{x\in X_i} T(X_\unsafe \mid x,\pi_{H-1}(x))\\
        &\leq  \max_{x\in X_i} \sum_{j=1}^{n_p+1} \gamma^{\pi}_{H,j} T(X_j\mid x,\pi_{H-1}(x))\\
        &= \gamma^{\pi}_{H-1}(x)
    \end{align*}
    For the induction case, assume $x\in X_\safe.$ Then, under the assumption that $V^{\pi}_{k+1}(x) \leq  \gamma^{\pi}_{k+1}(x)$ (induction hypothesis), it follows that:
    \begin{align*}
        V^{\pi}_{k}(x)& = \expect_{x'\sim T(\cdot \mid x,\pi_k(x))}[ V^{\pi}_{k+1}(x')  ] \\
        &\leq \expect_{x'\sim T(\cdot \mid x,\pi_k(x))}[ \gamma^{\pi}_{k+1}(x' ) ]\\
        &=\sum_{j=1}^{n_p+1} \gamma^{\pi}_{k+1,j} T(X_j\mid x,\pi_k(x))\\
        &=\gamma^{\pi}_k(x)  
    \end{align*}
\end{proof} 

\begin{remark}
    Note that if, in \eqref{eqn:StartingConditionValueIt}-\eqref{eqn:RecursiveConditionValueIt}, we replace the maximum operators with the minimum, then the solution of the resulting program returns a lower bound of $V^{\pi}_{0}$ and consequently an upper-bound on $P_\safe$.
\end{remark}


Since $\gamma_{0}^{\pi}$ is a piecewise-constant function and $\gamma_{0,i}^{\pi}$ is the unsafety probability bound for region $X_i$, \eqref{eqn:StartingConditionValueIt}-\eqref{eqn:RecursiveConditionValueIt} can be viewed as a DP for a discrete \emph{abstraction} of System~\eqref{eq:system_equation}.  In that view, the abstraction is a (time-inhomogenous) Markov chain (MC) $\M=(Q,P_\M)$, where $Q = \{q_1, \ldots, q_{n_p+1}\}$ is the state space such that $q_i$ represents region $X_i$, and $P_\M:Q\times Q\times \mathbb{N}_{<H} \to [0,1]$ is the transition probability function such that  
$P_\M(q_i,q_j,k)$ is the probability of transitioning from partition $X_i$ to $X_j$ at time step $k \in \naturals_{< H}$, i.e., 
\begin{equation*}
    P_\M(q_i,q_j,k) =
    \begin{cases}
        T(X_j \mid x^*, \pi(x^*)) & \text{if } i \neq n_p+1 \\
        1 & \text{if } i,j = n_p+1 \\
        0 & \text{otherwise,}
    \end{cases}
\end{equation*}
where $x^*=\arg\max_{x\in X_i}\sum_{j=1}^{n_p+1} \gamma^{\pi}_{k+1,j}  T(X_j\mid x,\pi_k(x))$. 

Computation of  $\gamma^{\pi}_k$ reduces to recursively solving the following maximization problem:
\begin{align}
    \label{Eqn:ValueIterationStepGamma}
    \max_{x \in X_i} \sum_{j=1}^{n_p+1} \gamma^{\pi}_{k+1,j}  T(X_j\mid x,\pi(x)).
\end{align}
That is, we need to find the input point that maximizes a linear combination of the transition probabilities for System~\eqref{eq:system_equation}. Since a positive weighted combination of convex (concave) functions is still convex (concave), \eqref{Eqn:ValueIterationStepGamma} can be solved exactly in the case  $T$ is either  concave or convex in $x$. 
Nevertheless, in the more general case, obtaining exact solutions of this optimization problem becomes infeasible. In what follows, we consider a relaxation of \eqref{Eqn:ValueIterationStepGamma} into a linear programming problem that leads to a well-known model, called interval Markov chains (IMC) or decision processes (IMDPs) {, which are a class of uncertain Markov processes where the exact transition probabilities between states are unknown but known to lie in some independent intervals \cite{givan2000bounded}. }

\subsection{Abstraction to Interval Markov Processes}
\label{sec:IMDPAbstraction}

To relax the maximization problem in \eqref{Eqn:ValueIterationStepGamma}, we note that  for each $x\in X_i$, $T(\cdot \mid x, \pi(x))$ is a discrete probability distribution over regions $X_1, \ldots, X_{n_p},X_\unsafe$. Consequently, \eqref{Eqn:ValueIterationStepGamma} weighs term $\gamma^{\pi}_{k+1,j}$ with the probability of transitioning to any of the $n_p + 1$ partitions of $X$.  This observation leads to 
Proposition~\ref{Prop:IMDP} below, where \eqref{Eqn:ValueIterationStepGamma} is relaxed to a linear programming problem. 
\begin{proposition}[{\cite{givan2000bounded}}]
\label{Prop:IMDP}
For partition $X_i\subseteq X_\safe,$ let $ {\bar{\gamma}^{\pi}_{k,i}}$ be the solution of the following linear programming problem:
\begin{align*}
   & {\bar{\gamma}^{\pi}_{k,i}}= \max_{t \in [0,1]^{n_p+1}} \sum_{j=1}^{n_p+1} \gamma^{\pi}_{k+1,j} \, t_j \\
   &\text{subject to:} \\
   & \qquad t_j \in \big[ \min_{x \in X_i}T( X_j \mid  x,\pi_k(x)), \max_{x \in X_i}T( X_j \mid  x,\pi_k(x)) \big], \\
   & \qquad \sum_{j=1}^{n_p+1} t_j =1,
\end{align*}
where $t_j$ is the j-th component of vector $t$.
Then, it holds that $ {\bar{\gamma}^{\pi}_{k,i}}\geq  {{\gamma}^{\pi}_{k,i}}=\max_{x \in X_i} \sum_{j=1}^{n_p+1} \gamma^{\pi}_{k+1,j} T( X_j \mid  x, \pi(x)) .$
\end{proposition}
 

In Proposition \ref{Prop:IMDP}, we model the likelihood of transitioning between each pair of states as intervals of probabilities, where the intervals are composed of all the valid transition probabilities for each point in the starting partition\edit{, i.e., each $t$ in the linear program in Proposition \ref{Prop:IMDP} represents a valid probability}. In this view, the introduced relaxation is in fact a relaxation of the MC abstraction $\M$ of System~\eqref{eq:system_equation} to an Interval~MC (IMC) \cite{nilim2005robust,givan2000bounded} $\mathcal{I} = (Q,\Plow_\I,\Pup_\I)$, where $Q$ is the same as in MC $\M$, and $\Plow_\I,\Pup_\I: Q\times Q \to [0,1]$ are, respectively, the lower and upper bound of the transition probability between each pair of states
such that
\begin{align*}
    &\Plow(q_i,q_j) = 
    \begin{cases}
        \min_{x \in X_i}T( X_j \mid  x,\pi_k(x)) & \text{if } i \neq n_p + 1 \\
        1 & \text{if } i,j = n_p+1 \\
        0 & \text{otherwise,}
    \end{cases}\\
    &\Pup(q_i,q_j) = 
    \begin{cases}
        \max_{x \in X_i}T( X_j \mid  x,\pi_k(x)) & \text{if } i \neq n_p + 1 \\
        1 & \text{if } i,j = n_p+1 \\
        0 & \text{otherwise.}
    \end{cases}
\end{align*}
Consequently, it follows that, in solving the LP in Proposition~\ref{Prop:IMDP}, we select the more conservative feasible distribution\footnote{A feasible distribution $t$ is a distribution that satisfies constraints of Proposition \ref{Prop:IMDP}.} with respect to probabilistic safety.

\begin{remark}
\label{Remark:EfficiencyIMDPIMplementation}
The optimization problem in Proposition \ref{Prop:IMDP} can be solved particularly efficiently due to the specific structure of the linear program. In particular, as shown in \cite{givan2000bounded}, one can simply order states based on the value of $\gamma_{k+1}^{\pi}$ and then assign upper or lower bounds based on the ordering and on the fact that $\sum_{j=1}^{n_p+1}t_j = 1.$  However, note that to solve the DP in \eqref{eqn:StartingConditionValueIt}-\eqref{eqn:RecursiveConditionValueIt}, the resulting linear program problem needs to be solved $H$ times for each of the $n_p+1$ states in $Q$.
\end{remark}

\begin{figure}[t]
    \centering
    \includegraphics[width=0.99\linewidth]{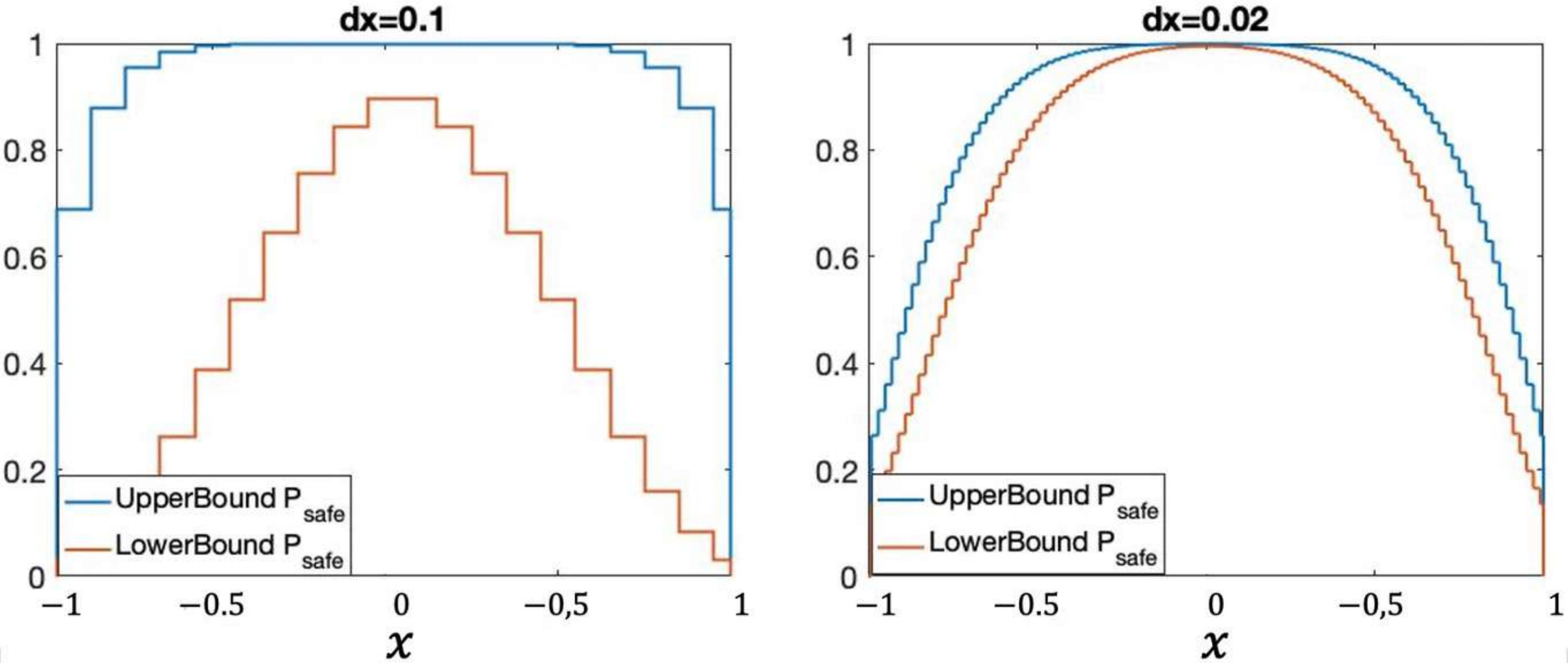}
    \caption{
    We consider the same setting as in \edit{Fig.} \ref{fig:ex_1_and_3} and for each state in the safe set, we plot upper and lower bounds of the probability of remaining within the safe set $[-1,1]$ for $10$ time steps starting from that state using two IMC abstractions: one obtained by discretizing the safe set uniformly with discretization step $dx=0.1$ and the other with $dx=0.02.$ For the initial set $X_0=[-0.25,0.25]$, we obtain a lower bound on $P_{\safe}$ of $0.756$ for the case $dx=0.1$ (left) and $0.975$ for $dx=0.02$ (right).
    }
    \label{fig:IMDP}
\end{figure}

\begin{remark}
Abstraction-based methods are numerical methods. As a consequence, their precision is dependent on the discretization of the state space. An example is illustrated in \edit{Fig.} \ref{fig:IMDP}, where we show how for a finer partition, the precision of approach increases (a proof of convergence to $P_s$ is given in Section \ref{th:CorrectnessControlIMDP}).  
How to optimally discretize the safe set, while limiting the explosion of the partition size, is still an active area of research, see e.g., \cite{adams2022formal,dutreix2018efficient}. \edit{We should also stress that the approach presented in this Section, while general, can sometimes be made tighter and less computationally expensive by adding to the optimization problem in Proposition~\ref{Prop:IMDP} additional constraints derived from additional assumptions on System~\eqref{eq:system_equation}. For instance, this is the case if $T$ has the product form, as shown in \cite{mathiesen2024scalable}. }
\end{remark}

\begin{remark}
    Note that an alternative approach for approximately solving \eqref{eqn:RecursiveConditionValueIt} would be to associate a representative point $x_i \in X_i $ to each partition and then consider the following approximation 
     \begin{multline}
        \label{Eqn:MarkovChainApproximatio}
        \max_{x \in X_i} \sum_{j=1}^{n_p+1} \gamma^{\pi}_{k+1,j}  T(X_j\mid x,\pi_k(x)) \approx \\
        \sum_{j=1}^{n_p+1} \gamma^{\pi}_{k+1,j}  T(X_j\mid x_i,\pi_k(x_i)).
    \end{multline} 
    The resulting abstraction leads to an MC with the same state space $Q$, and the transition probabilities between each pair of states are computed using representative points in the starting region as shown in \eqref{Eqn:MarkovChainApproximatio}.    {Such numerical approaches have been widely studied in the literature \cite{bertsekas1975convergence,abate2010approximate,lavaei2022automated}}, including also for continuous-time systems \cite{kushner2001numerical}, and approaches to quantify the resulting error have been developed \cite{abate2010approximate,abate2011approximate,esmaeil2015faust}. 
    \edit{Compared to the IMDP or IMC approaches described before, these approaches have the advantage to not require to solve any optimization problem over $x$. }
    However, similar to SBF (see Theorem \ref{th:StochasticBarrierFunction}), the resulting error generally  grows linearly with time \cite{abate2010approximate,lavaei2022automated} and is commonly more conservative compared to abstractions to IMC (or IMDP) \cite{cauchi2019efficiency}.
\end{remark}

\subsection{Control Synthesis with IMDPs}
\label{sec:ControlAbstractions}

For control policy computation, the set of controls or actions $U$ should be included in the abstraction model.  Hence, the abstraction becomes an IMDP $\I = (Q,U,\Plow,\Pup)$, where the transition probability bounds $\Plow,\Pup: Q \times U \times Q \to [0,1]$ are now also functions of $U$, i.e., given $q_i,q_j \in Q$ and $a \in U$, $\Plow(q_i,a,q_j)$ and $\Pup(q_i,a,q_j)$ are the lower- and upper-bound transition probability from $q_i$ to $q_j$ under action $a$, respectively.
An optimal policy via IMDP abstractions for region $X_i \subseteq X_s$ can then be computed by solving the following value iteration, which combines Proposition~\ref{Prop:IMDP} with \eqref{eqn:StartingConditionValueIt}-\eqref{eqn:RecursiveConditionValueIt}, and where we iteratively seek for the action that minimizes the probability that \edit{$\x_k$} enters the unsafe region $X_\unsafe$.
\begin{align}   
    & \label{eqn:ControlStartingConditionValueIt}\bar\gamma^{*}_{H,i}=\max_{x \in X_i} \ind{X_\unsafe}{x},\\
    \label{eqn:ControlRecursiveConditionValueIt}    
    & \bar\gamma^{*}_{k,i}=\min_{a\in U} \max_{t \in T^i_a}  \sum_{j=1}^{n_p+1} \bar\gamma^{*}_{k+1,j} t_j,
\end{align}
where for $i\leq n_p$ we define
\begin{multline*}
T^i_a=\Big\{ t\in [0,1]^{n_p+1} \mid 
 t_j \in \big[ \inf_{x \in X_i}T( X_j \mid  x,a),\\
 \sup_{x \in X_i}T( X_j \mid  x,a) \big], \; \sum_{j=1}^{n_p+1} t_j =1 \Big\}.
\end{multline*}
and
\begin{equation*}
    T^{n_p+1}_a=\{ t\in [0,1]^{n_p+1} \mid t_j=0 \text{ for $j\leq n_p$}, \; \, t_{n_p+1}=1  \}.
\end{equation*}
Note that in the inner maximization problem
in \eqref{eqn:ControlRecursiveConditionValueIt}, we seek a feasible distribution that maximizes the problem posed in Proposition~\ref{Prop:IMDP}. 
Consequently, synthesizing a policy according to \eqref{eqn:ControlStartingConditionValueIt}-\eqref{eqn:ControlRecursiveConditionValueIt} boils down to recursively solving the following min-max optimization problem for each partition $X_i \subseteq X_\safe$
\begin{align} 
\label{Eqn:OptimizationIMDP}
& \min_{a\in U} \max_{t \in T^i_a}  \sum_{j=1}^{n_p+1} \bar\gamma^{*}_{k+1,j} t_j .
\end{align}
If $U$ is discrete, then we simply need to apply Proposition \ref{Prop:IMDP} $|U|$ times, one for each action, and then take the action that minimizes the expression. 
If $U$ is uncountable, then in the case where $T$ is a convex or concave function of $a$,  a  solution to \eqref{Eqn:OptimizationIMDP} can be found efficiently via convex optimization \cite{delimpaltadakis2022interval}. For the more general cases, where one cannot rely on convex optimization, a sub-optimal policy can be found via heuristics or by discretizing $U$ \cite{dutreix2022abstraction}\edit{, but how to compute optimal policies for the general case still remains an open problem.}

The following theorem,  whose proof can be found in Section \ref{sec:proofs}, guarantees that the solution of the DP in \eqref{eqn:ControlStartingConditionValueIt}-\eqref{eqn:ControlRecursiveConditionValueIt} returns a lower bound on probabilistic safety for System~\eqref{eq:system_equation} and that the resulting policy converges to the optimal policy in the limit of a fine enough partition. 
\begin{theorem}
   \label{th:CorrectnessControlIMDP}
 Let $\bar\gamma^{*}_k$ be as defined in \eqref{eqn:ControlStartingConditionValueIt}-\eqref{eqn:ControlRecursiveConditionValueIt}. Then, for any $x\in X,$ it holds that 
 $ \bar{\gamma}^{*}_0(x)\geq V^*_0(x). $ 
 Furthermore, assume that  $X_\safe$ is a bounded hyper-rectangle, and that it is discretized uniformly. Then for every $ x\in X$, in the limit of $n_p\to\infty$, we have that $\bar{\gamma}^{*}_0(x) $ converges uniformly to $ V^*_0(x).$
\end{theorem}




\begin{remark}
    \label{Reark:ConvergenceRate}
Note that in the proof of Theorem \ref{th:CorrectnessControlIMDP}, reported in Section~\ref{sec:proofs}, we characterize the convergence rate of \eqref{eqn:ControlStartingConditionValueIt}-\eqref{eqn:ControlRecursiveConditionValueIt} to the optimal policy and to $P_s$. 
Nevertheless, the resulting convergence rate is generally conservative and to bound the error of abstraction-based methods, a much less conservative approach is as follows. For a given policy $\pi$, possibly obtained by solving \eqref{eqn:ControlStartingConditionValueIt}-\eqref{eqn:ControlRecursiveConditionValueIt}, solve  \eqref{eqn:StartingConditionValueIt}-\eqref{eqn:RecursiveConditionValueIt}, to get a lower bound of $P_\safe(X_\safe,x,H \mid \pi)$. Then, an upper bound can be similarly obtained by solving the following DP:
\begin{align}
    \label{eqn:ControlStartingConditionValueItSup}    
    & \gamma^{\pi,L}_H(x)=\gamma^{\pi,L}_{H,i}:=\min_{x \in X_i} \ind{X_\unsafe}{x},\\
    & \gamma^{\pi,L}_{k}(x)=\gamma^{\pi,L}_{k,i}:=  \min_{t \in T^i_{\pi_k(x)}} \sum_{j=1}^{n_p+1} \bar\gamma^{*}_{k+1,j} t_j,
    \label{eqn:ControlRecursiveConditionValueItSup} 
\end{align}
where $T^i_{\pi_k(x)}$ is defined as in \eqref{eqn:ControlRecursiveConditionValueIt}.    Intuitively, in the above problem, we iteratively seek a feasible distribution that minimizes the probability of reaching the unsafe region. Thus, $1-\gamma^{\pi,L}_{k}(x)$ returns an upper bound on $1-P_\safe(X_\safe,x,H \mid \pi)$, as illustrated in \edit{Fig.}~\ref{fig:IMDP}.
\end{remark}

%% file: Input/ProsandCons.tex

 {
Sections~\ref{sec:SBFs} and \ref{sec:abstractions} \edit{provide a unifying perspective on} SBF-based and abstraction-based methods under the framework of DP.  Specifically, these sections demonstrate that these methods are different approximation techniques for solving the same DP problem described in \eqref{Eq:OptimalValueIterationAction}-\eqref{eqn:valueIterationAction} (or equivalently in \eqref{Eq:FirstValueIteration}-\eqref{eqn:valueIteration}).
Hence, we can now fairly compare the pros and cons of the two approaches.  Namely, we focus on theoretical analysis with respect to the following properties: soundness, optimality, accuracy, computational effort, and scalability. 
}

\begin{itemize}

\item \textbf{Soundness:} As we show in Theorems \ref{th:StochasticBarrierFunction} and \ref{th:CorrectnessControlIMDP}, both approaches are guaranteed to return a valid lower bound on $P_{s}$.  Hence, both methods are sound.

\item \textbf{Optimality Guarantees:} As the $\beta$ term in Theorem \ref{th:StochasticBarrierFunction} is obtained by the supremum  expected change for all points in $X_\safe,$ the policy synthesized via SBF-based approaches is necessarily sub-optimal with a few exceptions 
 {
(e.g., 
when $\beta = 0$).
For this reason, there are no convergence guarantees of the safety certificate \edit{to $P_\safe$} for SBFs, in general.} 
In contrast, for IMDP-based approaches, as proved in Theorem~\ref{th:CorrectnessControlIMDP}, for a fine enough partition, both the policy and the certificate of safety converge to the optimal values for System~\eqref{eq:system_equation}.

\item \textbf{Accuracy:}  { As SBF-based approaches have no guarantees of convergence to the true $P_\safe$ (see discussion above)}, IMDP-based methods can generally provide less conservative bounds. However, 
  { as shown in Theorem \ref{th:CorrectnessControlIMDP},} this comes at the price of a fine enough partition.  {To illustrate this point, Table~\ref{tab:results} compares the bounds of $P_\safe$ obtained for the system in \edit{Fig.}~\ref{fig:ex_1_and_3}
 by using the state-of-the-art SBF methods and
 IMC abstractions of various sizes. 
 It is possible to observe that even barriers parameterized as neural networks or SoS polynomials of high degrees admit bounds that are far from the true probability even for this simple 1-dimensional linear example. This is in contrast with abstraction-based methods that converge to the true probability in the limit of arbitrarily large grid. 
 }

\begin{table}[t]
 {
\centering
\scalebox{1}{
\begin{tabular}{l l}
\toprule
Method &  $P_\safe$   \\
\midrule
SBF (SoS degree 2) \; $\cdots\cdots\cdots\cdots\cdots\cdots\cdots\cdots\cdots\cdots\cdots$  & $0.001$   \\
SBF (SoS degree 4) 
\; $\cdots\cdots\cdots\cdots\cdots\cdots\cdots\cdots\cdots\cdots\cdots$
& $0.8275$   \\
SBF (SoS degree 8) 
\; $\cdots\cdots\cdots\cdots\cdots\cdots\cdots\cdots\cdots\cdots\cdots$
& $0.8379$   \\
SBF (SoS degree 12)\, $\cdots\cdots\cdots\cdots\cdots\cdots\cdots\cdots\cdots\cdots\cdots$
& $0.8386$   \\
SBF (SoS degree 16)\, $\cdots\cdots\cdots\cdots\cdots\cdots\cdots\cdots\cdots\cdots\cdots$
& $0.8386$   \\
SBF (neural network 2 hidden layers with 32 neurons each)  & $0.802$   \\
SBF (indicator function) \ $\cdots\cdots\cdots\cdots\cdots\cdots\cdots\cdots\cdots\cdots$
& $0$   \\
\midrule
IMC ($dx=0.1$) 
\;\, $\cdots\cdots\cdots\cdots\cdots\cdots\cdots\cdots\cdots\cdots\cdots\cdots$
& $0.751$   \\
IMC ($dx=0.02$)  
\hspace{0.0mm}
$\cdots\cdots\cdots\cdots\cdots\cdots\cdots\cdots\cdots\cdots\cdots\cdots$
& $0.975$   \\
\midrule
True probabilistic safety & $\mathbf{0.9884}$   \\
\bottomrule
\end{tabular}
}
\caption{
\edit{
Comparison of lower bounds on $P_\safe$ obtained via various SBF-based and abstraction-based methods for the system in Fig.~\ref{fig:ex_1_and_3}: 
$\mathbf{x}[k+1]=\mathbf{x}[k] + \mathbf{v}$, with $\mathbf{v}\sim \mathcal{N}(0,0.1)$, safe set $X_s=[-1,1]$, initial set $X_0=[-0.25,0.25]$, and $H=10$. 
For the IMC methods, $dx$ is the width of the cells of a uniform grid that partitions $X_s$. For SBF in the case of SoS we use the approach presented in \cite{santoyo2021barrier} with various degrees of the SoS polynomial as reported in the table and with Lagrangian multipliers of degree $4$. In the case of neural network SBF, we used the approach developed in \cite{mathiesen2022safety} and ReLU activation functions. 
For the reference, true $P_\safe$ is also reported.
}
}
\label{tab:results}
}
\end{table}

\item \textbf{Computational Effort:} IMDP abstraction methods always require to partition the state space and to solve a DP over the resulting discretized space. Consequently, IMDP approaches necessitate to solve a number of linear programs that depends linearly on both the time horizon and number of states in the partition\edit{, which itself is exponential in the dimension of the state space of System~\eqref{eq:system_equation}.}
While, as explained in Remark~\ref{Remark:EfficiencyIMDPIMplementation}, each of  these problems can be solved efficiently and they are highly parallelizible, SBF-based approaches only require to solve a single (in general non-convex) optimization process. Furthermore, in the case of linear or polynomial dynamics, the resulting optimization problem can be often solved without any partitioning by employing existing tools from convex optimization.

\begin{table*}
    \begin{center}
    \begin{tabular}{  l  c  c   c   c   c   c   } 
      \toprule
      Approach & Soundness & Optimality & Accuracy & Computational Effort*    & Nonlinear Dynamics \\ 
      \midrule
      Stochastic Barrier Function &  $+$ & $-$ & $-$ & $+$  & $+$  \\ 
      \midrule
      Discrete Abstraction & $+$ & $+$ & $+$ & $-$  &  $+$\\
      \bottomrule
    \end{tabular}
    \caption{Summary of pros and cons of stochastic barrier function and discrete abstraction methods.  
     {*Computational Effort is with respect the the number of required optimization problems.}
    }
    \label{table:comparision}
    \end{center}
\end{table*}


\item \textbf{Scalability:}
In terms of scalability  {with respect to} the dimensionality of $X$,  abstraction-based methods face the state-space explosion problem since the size of discretization grows exponentially with dimensionality. \edit{Even if compositional approaches and methods based on model order reduction  have been developed to alleviate this problem \cite{lavaei2022constructing,lavaei2022automated}, still the issue remains. }  Similarly, as dimensionality grows, existing SBF methods face significant increase in complexity, depending on the parameterization of SBF. For instance, SoS-based approaches \cite{santoyo2021barrier} experience an exponential blow-up in the number of basis functions, e.g., number of monomials, leading to exponential growth in the optimization parameters.
Alternative approaches that parameterize an SBF as a neural network~\cite{mathiesen2022safety,vzikelic2023learning,abate2021learning} also experience exponential complexity in the dimension of $X$ to prove that a neural network is a valid SBF and to compute $\eta$ and $\beta$ as defined in Theorem~\ref{th:StochasticBarrierFunction}..
\edit{ Hence, scaling to high-dimensional systems remains a key challenge in safety verification of stochastic systems.}

\end{itemize}

 {A greatly-simplified summary of the discussion above is presented in Table~\ref{table:comparision}.  Note that the table reflects the theoretical analysis rather than empirical results. 
Also note that the computational effort comparison is based on the number of optimization problems that each method is required to solve.  As explained above, this is not necessarily equivalent to time complexity of the algorithms.}


From the above discussion, it becomes clear how abstraction-based methods and SBFs are complementary approaches, with a trade-off between computational demands and accuracy and/or flexibility in performing policy synthesis.  Consequently,  the choice of the method should depend on the particular application.

We should also mention that often one is interested in properties beyond safety and reachability, such as temporal logic properties. While this problem has been well studied for abstraction-based methods \cite{lahijanian2015formal}, \edit{where it has been shown to follow from the results for safety and reachability}, only few works have recently considered this problem for SBFs \cite{jagtap2020formal}.

%% file: Input/proofs.tex
\subsection{Proof of Theorem \ref{th:OptimalityMarkovPolicies}}
Define $P_{\unsafe}(X_\safe,x_0,H\mid \pi):=1-P_{\safe}(X_\safe,x_0,H\mid \pi).$ Then, to prove the result it is enough to show that 
$$\inf_{\pi \in \Pi}P_{\unsafe}(X_\safe,x_0,H\mid \pi)= V^{*}_0(x_0). $$
In the proof we are going to require the following lemma, which, for $\pi \in \Pi$ allows us to represent $P_{\unsafe}(X_\safe,x_0,H\mid \pi)$ via dynamic programming for history-dependent, possibly random, policies.
\begin{lemma}
\label{lemma:ValueFunctionExtended}
 For any $\pi=(\pi_0,...,\pi_{H-1}) \in \Pi,$ $k\in \{0,...,H\}$, let $\bar{V}_{k}^{\pi}:X^{k+1}\to [0,1]$ be defined as:
\begin{align*}
&\bar{V}^{\pi}_H(x_0,...,x_H)=\ind{X_\unsafe}{x_H}\\
    &\bar{V}_{k}^{\pi}(x_0,...,x_{k})=\ind{X_\unsafe}{x_{k}} +\\
    &\qquad \ind{X_\safe}{x_{k}} \expect_{x' \sim T(\cdot \mid x_{k}, \pi_k(x_0,...,x_{k}))}[ \bar{V}^{\pi}_{k+1}(x_0,...,x_{k},x')   ].
\end{align*}
Then, it holds that
 $$P_{\unsafe}(X_\safe,x_0,H\mid \pi)=\bar{V}^{\pi}_{0}(x_0).$$
\end{lemma}
\begin{proof}
The proof is by induction over the time index $k$. Define \begin{align*}
& P_{\unsafe}^{[x_0,...,x_k]}(X_\safe,[k,H]\mid \pi)= \\
& \quad P[\exists k'\in [k,H] \; s.t.\; \mathbf{x}_{k'}\in X_\unsafe\mid  \mathbf{x}_{0}=x_0,...,\mathbf{x}_{k}=x_k,\pi  ].
\end{align*}
Then, base case is for $k=H,$ that is  $P_{\unsafe}^{[x_0,...,x_H]}(X_\safe,[H,H]\mid \pi)=\bar{V}_{H}(x_0,...,x_H),$ which follows trivially. Then, assuming that $$P_{\unsafe}^{[x_0,...,x_{k+1}]}(X_\safe,[k+1,H]\mid \pi)=\bar{V}_{k+1}^{\pi}(x_0,...,x_{k+1}),$$ it holds that 
\begin{align*}
&    P_{\unsafe}^{[x_0,...,x_k]}(X_\safe,[k,H]\mid \pi)\\
=& P[ \exists k'\in [k,H] \; s.t.\; \mathbf{x}_{k'}\in X_\unsafe\mid  \mathbf{x}_{0}=x_0,...,\mathbf{x}_{k}=x_k,\pi ]\\
=& P[ \mathbf{x}_k \in X_\unsafe \vee \exists k'\in [k+1,H] \, s.t.\, \mathbf{x}_{k'}\in X_\unsafe  \\
&\hspace{4.8cm} \mid \mathbf{x}_{0}=x_0,...,\mathbf{x}_{k}=x_k,\pi ]\\
=&P[ \mathbf{x}_{k} \in X_\unsafe| \mathbf{x}_{k}=x_k ] \\
&+  P[ \exists k'\in [k+1,H] \, s.t.\, \mathbf{x}_{k'}\in X_\unsafe\mid  \mathbf{x}_{0}=x_0,...,\mathbf{x}_{k}=x_k,\pi ]\\
&- P[ \mathbf{x}_{k} \in X_\unsafe \wedge \exists k'\in [k+1,H] \, s.t.\, \mathbf{x}_{k'}\in X_\unsafe \\
&\hspace{4.8cm}\mid \mathbf{x}_{0}=x_0,...,\mathbf{x}_{k}=x_k,\pi]\\
=&\ind{X_\unsafe}{x_k}  +  P[ \exists k'\in [k+1,H] \, s.t.\, \mathbf{x}_{k'}\in X_\unsafe \\
&\hspace{4.8cm}\mid \mathbf{x}_{0}=x_0,...,\mathbf{x}_{k}=x_k,\pi ]\\
&- \ind{X_\unsafe}{x_k}P[ \exists k'\in [k+1,H] \, s.t.\, \mathbf{x}_{k'}\in X_\unsafe \\
&\hspace{4.8cm}\mid\mathbf{x}_{0}=x_0,...,\mathbf{x}_{k}=x_k,\pi]\\
=&\ind{X_\unsafe}{x_k}  +  \big(1- \ind{X_\unsafe}{x_k}\big)P[ \exists k'\in [k+1,H] \, s.t.\, \mathbf{x}_{k'}\in X_\unsafe \\
&\hspace{4.8cm} \mid\mathbf{x}_{0}=x_0,...,\mathbf{x}_{k}=x_k,\pi ]\\
=&\ind{X_\unsafe}{x_k}  +  \ind{X_\safe}{x_k} P[ \exists k'\in [k+1,H] \, s.t.\, \mathbf{x}_{k'}\in X_\unsafe \\
&\hspace{4.8cm} \mid\mathbf{x}_{0}=x_0,...,\mathbf{x}_{k}=x_k,\pi ]\\
=&\ind{X_\unsafe}{x_k}  +  \ind{X_\safe}{x_k}\int_{X} P[ \exists k'\in [k+1,H] \, s.t.\, \mathbf{x}_{k'}\in X_\unsafe \\
&\quad \mid \mathbf{x}_{0}=x_0,...,\mathbf{x}_{k}=x_k,\mathbf{x}_{k+1}=x',\pi ]P[\mathbf{x}_{k+1}=x'
\\
&\hspace{4cm}\mid\mathbf{x}_{0}=x_0,...,\mathbf{x}_{k}=x_k, \pi ] dx'\\
= & \ind{X_\unsafe}{x_k}  +  \ind{X_\safe}{x_k} \int_{X}\bar{V}_{k+1}^{\pi}(x_0,...,x_{k},x')P[\mathbf{x}_{k+1}=x'\\&\hspace{4cm}\mid\mathbf{x}_{0}=x_0,...,\mathbf{x}_{k}=x_k, \pi ] dx'\\
=&\bar{V}^{\pi}_{k}(x_0,...,x_k)
\end{align*}
\end{proof}
We can now prove the main statement. As $\Pi^{M,D}\subset \Pi$, it holds that $$\inf_{\pi \in \Pi} P_{\unsafe}(X_\safe,X_0,H\mid \pi) \leq \inf_{\pi \in \Pi^{M,D}} P_{\unsafe}(X_\safe,X_0,H\mid \pi).$$ Consequently, the proof is concluded if we can show that also holds that $\inf_{\pi \in \Pi} P_{\unsafe}(X_\safe,X_0,H\mid \pi) \geq \inf_{\pi \in \Pi^{M,D}} P_{\unsafe}(X_\safe,X_0,H\mid \pi).$
This can be done by induction over the time index $k\in \{0,H\}$. The base case is $k=H$, which follows trivially as $\forall x_0 \in X$ it holds that for any $\pi\in\Pi,$
$$ \bar{V}_{H}^{\pi}(x_0,...,x_H)={V}_{H}^{*}(x_H)=\ind{X_\unsafe}{x_H},$$
which is independent of the action and of the previous states.
Now assume that $\forall (x_0,...,x_{k+1}) \in X^{k+1}, \inf_{\pi \in \Pi}  V^{\pi}_{k+1}(x_0,...,x_{k+1})\geq V^*_{k+1}(x_{k+1})$, then it holds that
\begin{align*}
&\inf_{\pi \in \Pi}\bar{V}_{k}^{\pi}(x_0,...,x_k)\\
&=\inf_{\pi \in \Pi }\ind{X_\unsafe}{x_{k}} +  \\
&\qquad\quad \ind{X_\safe}{x_{k}}\expect_{x' \sim T(\cdot \mid x_{k}, \pi_k(x_0,...,x_{k}))}[ \bar{V}^{\pi}_{k+1}(x_0,...,x_{k},x')   ]  \\
&\geq \ind{X_\unsafe}{x_{k}} +  \\
& \qquad\qquad\quad\;\; \ind{X_\safe}{x_{k}} \inf_{\pi \in \Pi } \expect_{x' \sim T(\cdot \mid x_{k}, \pi_k(x_0,...,x_{k}))}[ {V}^{*}_{k+1}(x')   ] \\
&= \ind{X_\unsafe}{x_{k}} +  \ind{X_\safe}{x_{k}} \inf_{a \in U} \expect_{x' \sim T(\cdot \mid x_{k}, a)}[ V^{*}_{k+1}(x')   ]\\
 &= V^{*}_{k}(x_k)
\end{align*}

\subsection{Proof of Theorem \ref{th:CorrectnessControlIMDP}}

We recall that $\bar{\gamma}^{*}_{k}$ is the solution of the following dynamic programming problem
\begin{align*}   
    & \bar\gamma^{*}_H(x)=\bar\gamma^{*}_{H,i}:=\max_{x \in X_i} \ind{X_\unsafe}{x}\\   
    & \bar\gamma^{*}_{k}(x)=\bar\gamma^{*}_{k,i}:=\min_{a\in U} \max_{t \in T^i_a} \Big( \ind{X_\unsafe}{x}  + \\
    & \hspace{5cm} \ind{X_\safe}{x} \sum_{j=1}^{n_p+1} \bar\gamma^{*}_{k+1,j} t_j \Big),
\end{align*}
where for $i\leq n_p$
\begin{multline*}
    T^i_a=\Big\{ t\in [0,1]^{n_p+1} \mid t_j \in \big[ \inf_{x \in X_i}T( X_j \mid  x,a), \\
    \sup_{x \in X_i}T( X_j \mid  x,a) \big],\; \sum_{j=1}^{n_p+1} t_j =1 \Big\}
\end{multline*}
and for $i=n_p+1$ we have $T^{n_p+1}_a=\{ t\in [0,1]^{n_p+1} \mid t_j=0 \text{ for $j\leq n_p$} \, \wedge \, t_{n_p+1}=1  \}.$
Then, we first have to show that for any $k\in \{0,...,H\}$ and $x\in X$ in holds that 
 $ \bar\gamma^{*}_0(x)\geq V^*_0(x). $ This can be easily proved by contradiction. In fact,  assume that this statement is not true, then there must exist a strategy $\pi$ such that for $x\in X$ and $k\in \{0,...,H\}$ it holds that  $ \bar\gamma^{\pi}_k(x)\leq V^{\pi}_k(x). $ However, this contradicts Theorem \ref{th:CorrectnessGammaBoundsFixedControl} and Proposition \ref{Prop:IMDP}, thus concluding the proof that $ \bar\gamma^{*}_0(x)\geq V^*_0(x). $
A consequence of this  result is that for any $ \bar{x} \in X_i$, if we consider $a^*\in \argmin_{a\in U}  \expect_{x'\sim T(\cdot \mid \bar{x},a)}[V^*_{k+1}(x')],$ it holds that 
$$0 \leq \min_{a\in U}\max_{x\in X_i} \expect_{x'\sim T(\cdot \mid x,a)}[  \bar\gamma^*_{k+1}(x')]  -   \expect_{x'\sim T(\cdot \mid \bar x,a^*)}[V^*_{k+1}(x')]   $$
$$ \leq \max_{x\in X_i}\expect_{x'\sim T(\cdot \mid x,a^*)} [ \bar\gamma^*_{k+1}(x')]  -   \expect_{x'\sim T(\cdot \mid \bar x,a^*)}[V^*_{k+1}(x') ] .   $$
We will use this result to prove that for any $\epsilon >0$ there exists a partition of $X$ in $n_p+1$ regions such that 
 $ \forall x\in X, \, \bar\gamma^{*}_0(x) - V^*_0(x)\leq \epsilon. $ That, is $\bar{\gamma}^{*}_{0}(x)$ converges uniformly to $ V^*_0(x)$, thus concluding the proof.

Without any loss of generality, assume that $X_\safe$ is uniformly partitioned in $n_p$ hyper-cubes with the edge of size $\delta$. Further, consider an additional set in the partition $X_{n_p+1}=X_\unsafe.$ Then, the proof is as follows. We first show that for any $\bar\epsilon >0$ such that  $\max_{x\in X} \big( \bar\gamma^*_{k+1}(x)-V_{k+1}^*(x) \big) \leq \bar\epsilon$, then for all $x\in X$ it must hold that $\bar\gamma^*_{k}(x)-V_{k}^*(x)\leq \bar\epsilon+L\delta,$ where $L$ is such that for any $x,x'\in X_s$ and $X_j\subseteq X$ 
$$L|x-x'|_{\infty}\geq \max_{a\in U} |T(X_j\mid x,a)- T(X_j \mid x',a)|,  $$
which is guaranteed to exist because of the Lipschitz continuity of $T(X_j \mid x,a^*)$ wrt $x$.  Then, we derive a bound for $\bar\epsilon$ by considering the case $k=H-1$.
We start with the first part of the proof.


As mentioned, we start by assuming that $\forall x \in X, \bar{\gamma}^{*}_{k+1}(x)- V_{k+1}^{*}(x)\leq \bar{\epsilon}$. Under this assumption,  for $x\in X_i$ with $i\leq n_p$ select $a^*\in \argmin_{a\in U}  \expect_{x'\sim t(x'\mid x,a)}[V^*_{k+1}(x')],$  then the following holds
\begin{align*}
\bar{\gamma}^*_{k}(x) & - V_{k}^*(x)\\
\leq & \max_{t \in T^i_{a^*}} \sum_{j=1}^{n_p + 1} \bar{\gamma}^{*}_{k+1,j} t_j -\expect_{x'\sim T(\cdot \mid \bar x,a^*)}[V^*_{k+1}(x')] \\
\leq & \max_{t \in T^i_{a^*}} \sum_{j=1}^{n_p + 1} \bar{\gamma}^{*}_{k+1,j} t_j -\sum_{j=1}^{n_p + 1} \min_{x'\in X_j}V_{k+1}^{*}(x')  T(X_j\mid x,a^*)\\
\leq & \max_{t \in T^i_{a^*}} \sum_{j=1}^{n_p + 1} \big(\min_{x'\in X_j}V_{k+1}^{*}(x') + \bar{\epsilon} \big) t_j - \\
&\hspace{3cm}\sum_{j=1}^{n_p + 1} \min_{x'\in X_j}V_{k+1}^{*}(x')  T(X_j\mid x,a^*)\\
\leq & \bar{\epsilon} + \max_{t \in T^i_{a^*}} \sum_{j=1}^{n_p + 1} \min_{x'\in X_j}V_{k+1}^{*}(x')\big( t_j -  T(X_j\mid x,a^*) \big)\\
\leq & \bar{\epsilon} + \max_{t \in T^i_{a^*}} \max_{j\in \{1,...,n_p + 1\} } \big| t_j -  T(X_j\mid x,a^*) \big|\\
\leq & \bar{\epsilon} +  L\delta,
\end{align*}
where last two inequalities hold because $\sum_{j=1}^{n_p + 1} \min_{x'\in X_j}V_{k+1}^{*}(x')\leq \min_{x'\in X_j} \sum_{j=1}^{n_p + 1} V_{k+1}^{*}(x') \leq 1$ and becasue $t$ is a feasible distribution satisfying the constraints of Proposition \ref{Prop:IMDP}. Note that the case $x\in X_{n_p + 1} = X_u$ \edit{the inequality} follows trivially as in that case $\bar{\gamma}^*_{k}(x)=\ind{X_u}{x}=V_{k}^{*}(x).$ 
What is left to do in order to complete the proof is to obtain a bound of $\bar\epsilon$ as a function of $\delta.$ We can do that by noticing that 
\begin{multline*}
    \bar \gamma^{*}_{H-1}(x) - V_{H-1}^*(x) = \\
    \max_{t \in T^i_{a^*}} \big( t_{n_p+1} - T(X_\unsafe \mid x,a^* ) \big) \leq  L\delta.
\end{multline*}
Consequently,  to guarantee an error smaller than $\epsilon>0$, we can take $\delta \leq \frac{\epsilon}{HL},$ which results in $n_p\geq \big( \frac{lHL}{\epsilon}\big)^n + 1,$ where $l=\max_{x,x'\in X_s}|x-x'|_{\infty}$.